# Exposing Algorithmic Discrimination and Its Consequences in Modern Society: Insights from a Scoping Study

Ramandeep Singh Dehal
Cape Breton University
Sydney, NS, Canada
CBU22CLST@cbu.ca

Mehak Sharma
Cape Breton University
Sydney, NS, Canada
CBU22CLMP@cbu.ca

Ronnie de Souza Santos
University of Calgary
Calgary, AB, Canada
ronnie.desouzasantos@ucalgary.com

## ABSTRACT

*Algorithmic discrimination* is a condition that arises when data-driven software unfairly treats users based on attributes like ethnicity, race, gender, sexual orientation, religion, age, disability, or other personal characteristics. Nowadays, as machine learning gains popularity, cases of algorithmic discrimination are increasingly being reported in several contexts. This study delves into various studies published over the years reporting algorithmic discrimination. We aim to support software engineering researchers and practitioners in addressing this issue by discussing key characteristics of the problem.

**LAY ABSTRACT**. In recent years, data bias has posed significant challenges for minority groups, leading to systems and algorithms discriminating against Women, Black communities, people with disabilities, and LGBTQIA+ individuals. Recognizing this bias is the crucial first step toward devising solutions. This study maps and discusses instances of algorithmic discrimination across various societal domains.

## 1 INTRODUCTION

Modern society is diverse, multicultural, multifaceted, and highly dependent on computational solutions implemented under software engineering [1, 17]. Yet, despite the increasing emphasis on equity, diversity, and inclusion across many disciplines, software engineering still struggles with the challenges of producing software products that effectively represent this diversity [1]. Consequently, various types of data-driven algorithms and systems often discriminate against individuals from underrepresented groups [4, 12, 25].

Recently, researchers from various fields have highlighted instances of systematic bias in data-driven algorithms and systems that are causing discrimination. Examples include criminal justice algorithms disproportionately targeting Black individuals, and restricting their access to health and financial services [6, 14, 21]. In social media, digital filters have inadvertently altered skin tones and misidentified non-white individuals [15]. Another example is applications perpetuating stereotypes and limiting the experience of LGBTQIA+ individuals [23]. These cases represent just a fraction of the damage caused by biases originating in algorithms, which then reverberate through systems and affect the users.

Now, as with any other engineering, software engineering is concerned with the application of knowledge for the creation of wealth and quality of life in our society [24]. Hence, it is essential to thoroughly understand the harm algorithms inflict on individuals and utilize the reported problems to enhance software engineering methods, practices, and guidelines aiming to deliver inclusive and unbiased software; otherwise, wealth and quality of life will be restricted to some groups [1]. In this sense, this study provides a meta-synthesis of research from various disciplines highlighting the problems stemming from algorithmic discrimination. To address this topic, we conducted a scoping study aiming to answer the following research question: *What do we currently know about harms caused by algorithmic discrimination?*

From this introduction, our study is organized as follows. Section 2 explores concepts around algorithmic bias. In Section 3, we outline our methodology, while Section 4 presents the findings from our meta-synthesis. Section 5 discusses our findings and presents the implications of our study. Lastly, Section 6 summarizes our contributions.

## 2 BACKGROUND

Algorithmic discrimination is a condition in which data-driven software contributes to unfair treatment of users based on their background, which includes attributes like ethnicity, race, gender, sexual orientation, religion, age, national origin, disability, or any other personal characteristic [12, 18]. This problem is intricately tied to biases existing in the data used within the system, as these biases can sustain or amplify preconceived prejudices, leading to discrimination, with a particular impact on individuals from underrepresented groups [13].

Biases in data-driven systems are particularly worrisome when they influence critical social decisions, such as determining eligibility for release from incarceration, access to loans, allocation of healthcare resources, job hiring decisions or experiences on the internet [2, 4, 10, 12, 16, 19, 25]. For this reason, algorithmic discrimination is recognized as a collective concern that transcends technology since it affects not only computer science and software engineering but also disciplines such as law, sociology, health, ethics, and beyond [9].

Currently, within the domain of software engineering, significant efforts are underway to tackle algorithmic discrimination. These initiatives approach the problem from various angles, including the improvement of software development [3, 20], the implementation of fairness testing to mitigate bias [7], and considerations from a software management standpoint [20, 25]. However, while software engineering has yet to make significant progress in addressing this issue, other disciplines keep systematically reporting algorithmic discrimination problems and demanding solutions from those involved in technology development [21].

## 3 METHOD

Secondary studies are a research approach frequently used to consolidate findings from various primary studies, including experiments, case studies, surveys, and experience reports, among others [11, 26]. This study is a scoping review [22] designed to identify and explore papers and articles published over the years that documented and reported issues arising from algorithmic discrimination. We aim to construct an overview of this theme, which can guide researchers in exploring and addressing questions related to this problem. To accomplish this, we followed the guidelines for conducting systematic reviews in software engineering [11], as detailed below.

### 3.1 Specific Research Questions

As our primary goal is to provide an overview of research on algorithmic discrimination drawing from studies published across various disciplines, we formulated five specific research questions that guided our collection, analysis, and synthesis of evidence: *RQ1.* How has the number of studies reporting algorithmic discrimination evolved over the years? *RQ2.* Which disciplines are reporting issues and problems related to algorithmic discrimination? *RQ3.* What types of discrimination caused by algorithms are documented in the literature? *RQ4.* What problems resulting from algorithmic discrimination are reported in the literature?

### 3.2 Data Sources and Search Strategy

We conducted a semi-automated search process by applying a search string across four search engines and indexing systems and then, manually selecting studies that focused on the topic under investigation. This search string was constructed using terms extracted from our general research question: *algorithm*, *discrimination*, and *bias*. In this process, we referred to IEEE Xplore, ACM Digital, Science Direct, and Google Scholar to identify evidence about the problem.

We opted for a semi-automated approach because search engines differed on how they structure search strings. In addition, many initially retrieved papers did not align with the scope of our research, which specifically targeted studies **reporting the harms caused by algorithmic discrimination** and not simply mentioning it. The automated search was concluded in the second half of 2023 and retrieved 305 papers potentially relevant to our analysis. Subsequently, after applying exclusion and inclusion criteria, we reduced this number to 97 papers.

### 3.3 Inclusion and Exclusion Criteria

We applied five exclusion criteria and one inclusion criterion to evaluate the 305 studies retrieved through our search. First, papers were excluded from our research if they met any of the following exclusion criteria: a) *EC-1* Not written in English; b) *EC-2* Unavailable for download or inaccessible for online reading; c) *EC-3* Incomplete texts, like drafts, presentation slides, or abstract only; d) *EC-4* Proposing tools for a particular scenario without offering a thorough discussion of the problem. After applying the exclusion criteria, we specifically focused on selecting papers that addressed problems associated with algorithmic discrimination, and, additionally, those that provided insights, lessons learned, or experience reports.

### 3.4 Selection Process

Initially, papers were identified using the general search string. Subsequently, we manually narrowed down the list of retrieved papers by selecting those that aligned with the scope of this study based on the analysis of the following metadata: title, abstract, and keywords. Following this, we applied the set of exclusion criteria. Finally, the inclusion criterion was employed to establish the definitive set of papers that underwent comprehensive analysis in our study. Many papers were excluded during the exclusion/inclusion criteria phase as they did not present substantial evidence on the investigated topic. Instead, they merely referenced the topic or provided vague information without discussing the real-world implications of algorithmic discrimination.

### 3.5 Extraction, Analysis and Synthesis

We implemented a structured form to facilitate the data extraction process. During this stage, we accessed the full text of each paper and recorded the relevant information to answer our specific research questions (Section 3.1). To enhance the accuracy of data extraction and thereby improve the reliability of our results, two researchers collaborated on this process. Any discrepancies in extraction were resolved through discussion and consensus, and a third researcher was involved when necessary. After completing the form with the evidence gathered from the papers, we employed two data analysis approaches to synthesize the obtained data and present our findings. We started by applying descriptive statistics [8] to summarize the distribution and frequency of papers by year, discipline, and the type of discrimination reported in the studies. Furthermore, we conducted a thematic analysis [5] to explore the evidence presented in the papers in-depth and derive implications for software engineering.

## 4 FINDINGS

A total of 97 unique papers were identified and analyzed in this scoping study. The complete list of papers is accessible https://figshare.com/s/a57b4a572abc5e880abb. Below, we address each of the specific research questions with the evidence collected in this study.

### 4.1 RQ1. Evolution of Studies on Algorithmic Discrimination

Over the last decade, there has been a noticeable increase in the number of studies uncovering instances of algorithmic discrimination. The first study on this topic identified in our research dates back to 2009. Starting in 2013, there has been a gradual rise in the annual publication of papers, culminating in a historic peak in 2022. It is important to acknowledge that our data collection occurred before the conclusion of 2023, leaving open the possibility of additional papers being published during the current year. Figure 1 illustrates the evolution of publications between 2019 and the first half of 2023. The year 2018 marked a significant turning point, as there was a substantial increase in the number of studies about algorithmic discrimination, signaling a growing focus on this issue within academic communities. This surge continued to escalate in the following years, with 36 studies documented in 2019 and 42 in 2022. These numbers can be seen as a reflection of the growing volume of research in the field of machine learning and the widespread adoption of artificial intelligence-based software and applications in our society, which, in turn, led to an increase in reported issues associated with these algorithms. Furthermore, this is the period when conferences and workshops dedicated to fairness and responsible AI began to emerge, fostering increased discussions on the topic.

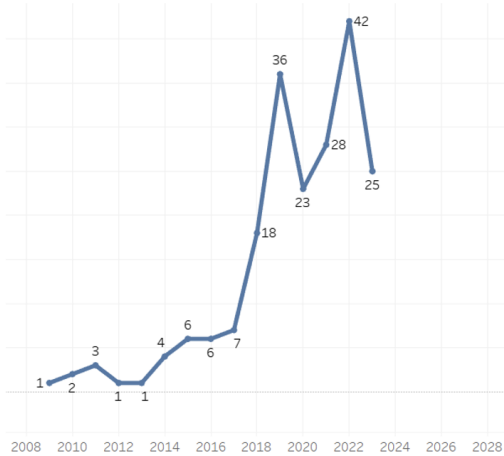

**Figure 1: Evolution of Publications Over the Years**

## 4.2   RQ2. Disciplines Reporting Algorithmic Discrimination

Our review identified four research domains engaged in discussions on algorithmic discrimination. Taking a broader perspective, these areas can be divided into two primary groups: a) Authors from disciplines closely associated with software development, such as Computer Science and Software Engineering; b) Authors from disciplines that primarily consume software and technologies, including Social Sciences, Law, and Healthcare. We observed an overlap in these papers resulting from the collaboration among authors from different fields, e.g., a paper about algorithmic discrimination published by researchers from software engineering and health.

A clear distinction emerges between these two groups. In the *technology-developers group*, studies cover a wide spectrum of topics related to algorithmic discrimination, ranging from recognizing the issue as a gap in AI applications to proposing mitigation strategies and identifying areas for further research. Conversely, the *technology-users group* is more direct in highlighting algorithmic problems and their negative societal consequences, frequently calling for increased attention and responsibility from technology fields in tackling this issue.

Outside the field of software engineering, studies provide a comprehensive exploration of the consequences of algorithmic discrimination within our society. This is particularly evident when considering the problems these algorithms, especially decision-making algorithms, have caused for individuals from underrepresented groups. We delve into these issues in greater detail in Section 5.1.

## 4.3   RQ3. Types of Algorithmic Discrimination

We identified seven types of discrimination caused by algorithms that were reported over the years. Our findings highlight that gender and ethnicity discrimination are the most frequently discussed topics in the studies, with 83 and 58 papers. Additionally, ageism was identified in 20 papers. Prejudice based on culture and disabilities is also notable, with 16 and 15 papers, respectively, reporting these problems. In contrast, discrimination stemming from biases related to sexual orientation and geographical location received less attention, with seven and four papers, respectively, reporting these specific issues.

## 4.4   RQ4. Algorithmic Discrimination Issues

Our review demonstrates that algorithmic discrimination is the source of numerous challenges that currently affect both individuals and society, including unfair treatment of users, privacy concerns, economic issues, and beyond. Below, Table 1 presents the main issues reported in the literature.

## 5   DISCUSSION

We start this section with an overall discussion on algorithmic discrimination derived from our meta-synthesis. Subsequently, we outline the limitations of our study.

### 5.1   Exposing Algorithmic Discrimination

Algorithmic discrimination is a complex problem with far-reaching consequences in modern society, influencing work, access to services, ethical considerations, and social outcomes. Biased algorithms developed in software engineering contribute to inequalities and might harm marginalized groups across diverse domains. Consequently, this problem has prompted discussions and claims for solutions from disciplines beyond software engineering.

In this context, diverging from the common narrative found in mainstream sources, including non-academic platforms like forums and social media, where algorithmic discrimination is primarily linked to gender and ethnicity concerns, our meta-review has identified seven distinct types of discrimination experienced by users as a result of biases within data-driven systems. These findings underscore that algorithmic discrimination arises from biases associated with individual characteristics and sensitive user data, which request a multifaceted mitigation approach.

This scenario entails that, to effectively address this issue, strategies must include identifying and mitigating various bias types, adapting solutions to diverse contexts, implementing rigorous processes to evaluate data sources, adhering to ethical principles and

Table 1: Problems Caused by Algorithmic Discrimination

| Discrimination | Issue | Papers |
| --- | --- | --- |
| Gender | Discrimination against women, a persistent societal issue, gains prominence when linked with bias in data-driven applications. Gender wage disparities persist, influenced by clustering algorithms. Advertising platforms exhibit gender-based ads, influencing how job opportunities are perceived. Search engines, although not explicitly focused on gender, indirectly contribute to inequalities and reinforce stereotypes against women. Additionally, classification, data mining, and prediction algorithms focus primarily on binary gender distinctions, which raises concerns for non-binary individuals. In healthcare, for instance, non-binary gender considerations are often overlooked in data-driven systems, posing risks of patient misdiagnosis and unfairness. | P001, P002, P005, P007, P008, P009, P010, P011, P012, P014, P017, P020, P021, P022, P023, P024, P025, P026, P027, P028, P029, P030, P032, P034, P035, P037, P038, P039, P040, P041, P042, P043, P044, P045, P046, P047, P048, P049, P050, P051, P052, P053, P054, P055, P056, P057, P058, P059, P060, P061, P062, P063, P064, P065, P066, P067, P068, P069, P070, P071, P072, P073, P074, P075, P076, P077, P078, P079, P080, P083, P084, P086, P087, P088, P089, P090, P091, P092, P093, P094, P095, P096, P097 |
| Ethnicity | Instances of racism are increasing within data-driven systems, leading to significant disparities, particularly for underrepresented ethnic groups, in key areas like job opportunities and access to economic and healthcare resources. Algorithmic racism is also evident in biased search results, particularly images on social media and ad libraries, which have been reported to discriminate against Black and Hispanic users. | P003, P005, P010, P015, P016, P017, P019, P020, P023, P024, P025, P026, P028, P031, P034, P035, P038, P040, P041, P043, P044, P049, P050, P051, P052, P053, P054, P055, P056, P058, P060, P061, P063, P064, P065, P066, P067, P068, P070, P071, P072, P073, P074, P075, P076, P077, P079, P081, P083, P084, P085, P087, P090, P092, P093, P094, P095, P096 |
| Age | In a period characterized by AI-driven automated decision-making, ageism can become a notable concern. As an example of issues recently reported, credit assessments may unintentionally discriminate against older individuals, and skewed training data in facial analysis algorithms can introduce challenges in pattern recognition affecting various services. | P009, P020, P021, P028, P034, P038, P041, P043, P046, P058, P062, P066, P067, P068, P076, P077, P079, P084, P090, P095 |
| Culture | Presently, there is a growing concern about data-driven systems potentially perpetuating discrimination rooted in cultural factors such as religion, political beliefs, and socioeconomic status. The incorporation of these attributes in predictive analysis has the potential to unfairly assign failure or trigger exclusion to individuals in various contexts, including educational predictions and analytics. | P010, P013, P020, P023, P031, P035, P043, P052, P060, P065, P066, P073, P074, P081, P090, P094 |
| Disability | Biased data-driven software has the potential to stigmatize and aggravate the challenges faced by individuals with disabilities. Some AI tools have enabled corporations to collect and analyze user data, and potentially assign disabilities based on online interactions. Additionally, discrimination against people with disabilities has been reported across various domains, including employment, education, public safety, and healthcare, affecting those with both physical and cognitive impairments. | P004, P006, P017, P018, P019, P033, P035, P036, P058, P059, P062, P065, P069, P082, P090 |
| Sexual Orientation | Algorithmic discrimination related to sexual orientation, and commonly manifested in real-time and online platforms. Instances of discrimination can be found when decision-making processes restrict the access of LGBTQIA+ individuals to various activities or censor their participation in collective online spaces, thus perpetuating social stigmatization. | P061, P063, P065, P086, P089, P090, P097 |
| Geography | Discrimination in data-driven algorithms can extend beyond personal cultural characteristics and create prejudice based on people's geographic location. This problem can lead to inequalities in resource access, particularly during crises like wars and natural disasters, where aid distribution may be influenced by biased geographic data, that has the potential to intensify pre-existing disparities and contribute to restricting or completely impeding humanitarian initiatives. | P032, P035, P057, P058 |

Papers available on https://figshare.com/s/a57b4a572abc5e880abb

legal regulations, engaging users for valuable feedback, promoting interdisciplinary collaboration, and combating evolving biases. In summary, given the pervasive nature of algorithmic discrimination, developing unbiased data-driven systems requires a primary focus on deploying comprehensive and adaptable approaches.

Lastly, despite ongoing efforts to address this issue, advancements in software engineering appear insufficient in light of numerous cases being constantly reported. Our meta-review emphasizes the importance of directly tackling algorithmic bias and actively advocating for fairness and inclusiveness within machine learning and software development to effectively combat algorithmic discrimination against individuals.

## 5.2 Limitations

In our study, we acknowledge limitations in our search strategy since two out of four repositories used to identify papers are primarily focused on computer science and software engineering research. Hence, while we obtained a substantial number of relevant papers, we acknowledge that additional examples of algorithmic discrimination might be reported in papers from diverse disciplines available in different repositories—identifying these papers is part of our plans for future research. Finally, we reduced selection and extraction biases by employing a dual-reviewer process with consensus meetings and additional third-reviews as needed.

## 6 CONCLUSION

As the widespread adoption of data-driven systems supported by machine learning experiences grows fast, concerns regarding data bias have become increasingly evident across various domains. Within this context, the rise of algorithmic discrimination emerges as a contemporary issue, particularly impacting individuals belonging to underrepresented groups in our society. As software engineering gradually delves into addressing this gap, other fields highlight the eminent risks of algorithmic discrimination and advocate for more robust measures to mitigate the increasing frequency of this issue. In our scoping study, we identified 97 papers outlining seven types of discrimination resulting from algorithmic bias: gender, ethnicity, age, culture, disability, sexual orientation, and geography.

In summary, the pervasive occurrence of algorithmic discrimination underscores the need to prioritize the implementation of a diverse range of adaptable solutions. We believe that our findings can assist software engineering researchers in recognizing the problem and identifying opportunities to address the gap. Regarding our future work, our initial plan is to extend this meta-review by incorporating experiences reported in other scientific repositories and in the grey literature, which is expected to enrich the body of knowledge about the problem. Our long-term research goal is to assist software professionals in confronting algorithmic discrimination by delving into not only the technical facets of software but also the human aspects of software development, such as team diversity, which has been recognized as essential for fostering the development of inclusive technologies.

## REFERENCES
[1] Khaled Albusays, Pernille Bjorn, Laura Dabbish, Denae Ford, Emerson Murphy-Hill, Alexander Serebrenik, and Margaret-Anne Storey. 2021. The diversity crisis

in software development. *IEEE Software* 38, 2 (2021), 19–25.
[2] Robin Allen and Dee Masters. 2020. Artificial Intelligence: the right to protection from discrimination caused by algorithms, machine learning and automated decision-making. In *ERA Forum*, Vol. 20. Springer, 585–598.
[3] Rachel KE Bellamy, Kuntal Dey, Michael Hind, Samuel C Hoffman, Stephanie Houde, Kalapriya Kannan, Pranay Lohia, Jacquelyn Martino, Sameep Mehta, Aleksandra Mojsilović, et al. 2019. AI Fairness 360: An extensible toolkit for detecting and mitigating algorithmic bias. *IBM Journal of Research and Development* 63, 4/5 (2019), 4–1.
[4] Yuriy Brun and Alexandra Meliou. 2018. Software fairness. In *Proceedings of the 2018 26th ACM joint meeting on european software engineering conference and symposium on the foundations of software engineering*. 754–759.
[5] Daniela S Cruzes and Tore Dyba. 2011. Recommended steps for thematic synthesis in software engineering. In *2011 international symposium on empirical software engineering and measurement*. IEEE, 275–284.
[6] Jane E Fountain. 2022. The moon, the ghetto and artificial intelligence: Reducing systemic racism in computational algorithms. *Government Information Quarterly* 39, 2 (2022), 101645.
[7] Sainyam Galhotra, Yuriy Brun, and Alexandra Meliou. 2017. Fairness testing: testing software for discrimination. In *Proceedings of the 2017 11th Joint meeting on foundations of software engineering*. 498–510.
[8] Darren George and Paul Mallery. 2018. Descriptive statistics. In *IBM SPSS Statistics 25 Step by Step*. Routledge, 126–134.
[9] Bryce W Goodman. 2016. Economic models of (algorithmic) discrimination. In *29th conference on neural information processing systems*, Vol. 6.
[10] Pauline T Kim. 2021. Addressing algorithmic discrimination. *Commun. ACM* 65, 1 (2021), 25–27.
[11] Barbara A Kitchenham, Tore Dyba, and Magne Jorgensen. 2004. Evidence-based software engineering. In *Proceedings. 26th International Conference on Software Engineering*. IEEE, 273–281.
[12] Jon Kleinberg, Jens Ludwig, Sendhil Mullainathan, and Cass R Sunstein. 2018. Discrimination in the Age of Algorithms. *Journal of Legal Analysis* 10 (2018), 113–174.
[13] Nima Kordzadeh and Maryam Ghasemaghaei. 2022. Algorithmic bias: review, synthesis, and future research directions. *European Journal of Information Systems* 31, 3 (2022), 388–409.
[14] Nicol Turner Lee. 2018. Detecting racial bias in algorithms and machine learning. *Journal of Information, Communication and Ethics in Society* 16, 3 (2018), 252–260.
[15] Michael Lutz, Sanjana Gadaginmath, Natraj Vairavan, and Phil Mui. 2021. Examining political bias within youtube search and recommendation algorithms. In *2021 IEEE Symposium Series on Computational Intelligence (SSCI)*. IEEE, 1–7.
[16] Mason Marks. 2019. Algorithmic disability discrimination. *Disability, Health, Law and Bioethics (Cambridge University Press, 2020)* (2019).
[17] Aasif Ali Naikoo, Shashank Shekhar Thakur, Tariq Ahmad Guroo, and Aadil Altaf Lone. 2018. Development of society under the modern technology-a review. *Scholedge International Journal of Business Policy & Governance* 5, 1 (2018), 1–8.
[18] Safiya Umoja Noble. 2018. Algorithms of oppression. In *Algorithms of oppression*. New York university press.
[19] Natalia Norori, Qiyang Hu, Florence Marcelle Aellen, Francesca Dalia Faraci, and Athina Tzovara. 2021. Addressing bias in big data and AI for health care: A call for open science. *Patterns* 2, 10 (2021).
[20] Kalia Orphanou, Jahna Otterbacher, Styliani Kleanthous, Khuyagbaatar Batsuren, Fausto Giunchiglia, Veronika Bogina, Avital Shulner Tal, Alan Hartman, and Tsvi Kuflik. 2022. Mitigating Bias in Algorithmic Systems—A Fish-Eye View. *Comput. Surveys* 55, 5 (2022), 1–37.
[21] Kellie Owens and Alexis Walker. 2020. Those designing healthcare algorithms must become actively anti-racist. *Nature medicine* 26, 9 (2020), 1327–1328.
[22] Paul Ralph and Sebastian Baltes. 2022. Paving the way for mature secondary research: the seven types of literature review. In *Proceedings of the 30th ACM Joint European Software Engineering Conference and Symposium on the Foundations of Software Engineering*. 1632–1636.
[23] Julian A Rodriguez. 2023. LGBTQ incorporated: YouTube and the management of diversity. *Journal of Homosexuality* 70, 9 (2023), 1807–1828.
[24] Guenther Ruhe, Maleknaz Nayebi, and Christof Ebert. 2017. The vision: Requirements engineering in society. In *2017 IEEE 25th International Requirements Engineering Conference (RE)*. IEEE, 478–479.
[25] Ronnie de Souza Santos, Luiz Fernando de Lima, and Cleyton Magalhaes. 2023. The Perspective of Software Professionals on Algorithmic Racism. (2023).
[26] Ronnie ES Santos, Cleyton VC de Magalhães, and Fabio QB da Silva. 2014. The use of systematic reviews in evidence based software engineering: a systematic mapping study. In *Proceedings of the 8th ACM/IEEE international symposium on empirical software engineering and measurement*. 1–4.